# All-optical switching due to state-filling in quantum dots

*R. Prasanth, J.E.M. Haverkort, A. Deepthy, E.W. Bogaart, J.J.G.M. van der Tol, E.A. Patent, G. Zhao, Q. Gong, P.J. van Veldhoven, R. Nötzel and J.H. Wolter*

*eiTT/COBRA Inter-University Research Institute, Eindhoven University of Technology, P.O. Box 513, 5600MB Eindhoven, The Netherlands*

**Abstract:** We report all-optical switching due to state-filling in quantum dots (QDs) within a Mach-Zehnder Interferometric (MZI) switch. The MZI was fabricated using InGaAsP/InP waveguides containing a single layer of InAs/InP QDs. A 1530-1570 nm probe beam is switched by optical excitation of one MZI-arm. By exciting below the InGaAsP bandgap, we prove that the refractive index nonlinearity is entirely due to the QDs. The switching efficiency is 5 rad/(µW absorbed power). Probe wavelength insensitivity was obtained using a broad size distribution of QDs.

All-optical switching has mainly been performed using active elements such as semiconductor optical amplifier gates[1,2]. Photonic switching in passive materials[3-6] suffer from small all-optical nonlinearities, requiring a too high switching energy. Semiconductor quantum dots (QDs) are expected to provide improved all-optical nonlinearities[7,8] due to their delta-function density of states. QDs show sharp excitonic peaks with considerably larger peak absorption than in bulk or quantum wells. We recently calculated a 35x enhancement of the electrorefraction in homogeneous QDs[8] as compared to quantum wells. We expect a similar enhancement for the all-optical refractive index nonlinearity due to state-filling in QDs. The refractive index nonlinearity in QDs is also enhanced, since a single electron-hole pair is able to induce transparency of the ground state transition, while 2 electron-hole pairs generate optical gain. In this letter we investigate all-optical switching in the 1550 nm wavelength window due to state-filling of a single layer of InAs/InP QDs embedded in a InGaAsP/InP waveguide, processed into a Mach-Zehnder Interferometric (MZI) space switch.

The all-optical switching set-up is schematically shown in Fig. 1. The pump beam is provided by a tunable optical parametric oscillator (OPO) which delivers 200 fs pulses at 76 MHz repetition rate. The pump beam excites one of the two arms of the MZI from the top, i.e. perpendicular to the substrate. The OPO ($\lambda$>1400 nm) excites carriers directly into the QDs without exciting the bulk InGaAsP or InP. The resulting state-filling in the InAs/InP QDs produces a refractive index variation leading to switching of the MZI. The switching of the MZI is probed with a CW tunable semiconductor laser, tuned into the 1530-1570 nm wavelength window. The probe beam is coupled into the MZI by microscope objectives.

The probe output is focussed onto a slit to spatially separate the two outputs of the MZI. The all-optical switching signal, as shown in Fig. 2, is acquired by chopping (2 kHz) the pump beam and measuring the demodulated probe output with a lock-in amplifier. The pump laser excites a surface area of approximately 600x25 μm$^2$ around the upper arm of the MZI, as schematically indicated in Fig. 1. Since we pump with a mode-locked OPO, generating carriers with a repetition rate of 76 MHz and a typical decay time[3,9] of 55-65 ps, while we probe CW, we measure a time-averaged switching efficiency, estimated to be 0.5% of the peak switching efficiency. Great care has been exercised to avoid spurious contributions due to photoluminescence (PL) guided within the waveguide as well as unwanted thermal switching. The magnitude of the PL-contribution was regularly checked by fine-tuning the semiconductor probe laser, thereby separating the oscillating interferometric switching signal from the constant PL background. Thermally activated switching is not expected for excitation energies below the InGaAsP bandgap where only 0.08 % of the pump light is absorbed[9] by the QDs, resulting in a heating power of 10 nW directly hitting the waveguide. The resulting $\Delta n^{thermal} \leq 10^{-7}$ is negligible compared to the observed refractive index nonlinearity.

The QD sample was grown by Chemical Beam Epitaxy on a (100) oriented InP substrate. The QDs are prepared by Stranski Krastanow growth by depositing 4.3 monolayers InAs at 500$^o$C on top of a lattice matched $Ga_xIn_{1-x}As_yP_{1-y}$ layer (x=0.2515, y=0.546). The QDs are subsequently capped with 0.62 nm InP and annealed for 5 min in PH$_3$, shifting the PL peak wavelength 100 nm down to 1500nm. Atomic Force Microscopy of similar uncapped InAs/InP QDs shows a density of $1.4.10^{10}/cm^2$. The single QD layer is embedded into a 370 nm thick Q1.3 InGaAsP waveguide core, which is covered with a 1.3 µm InP cladding.

Subsequently, 2x2 Mach-Zehnder Interferometric space switches[10,11] were fabricated, built on 3-dB Multi-Mode Interference (MMI) input and output couplers. The structures were defined in 100 nm $SiN_x$. The ridge waveguides were shallow-etched with a depth of 30 nm into the waveguide core using a $CH_4/H_2$ RIE process and an $O_2$ descumming process. The upper and lower arms of the MZI have a width of 2.8 and 3.55 µm respectively. The length of the phase shifting section is 605 µm, with 30 µm separation between the arms.

Room temperature PL shown in the inset of Fig. 3, reveals an InGaAsP peak at 1300 nm with 50 meV FWHM and a QD PL peak between 1400 and 1600 nm with 90 meV FWHM. The QD size distribution was kept broad for obtaining a wavelength insensitive switching behaviour. The waveguide loss is 60 dB/cm for TE and 11 dB/cm for TM-polarization, allowing photonic switching experiments with a 0.3 mW TM-polarized probe. Due to the large waveguide loss, Fabry-Perot effects due to reflections between the chip facets are small compared to the all-optical switching signal.

Fig.2 shows the all-optical switching results for excitation of the QDs at 1450 nm and detection between 1530 and 1570 nm. The demodulated probe transmission has opposite sign for the two outputs of the MZI, indicating all-optical switching. The pump laser excitation density of 1W/$cm^2$ corresponds to a relative QD occupation of 1.4% at the highest power of 0.125 mW presented in Fig. 2. The 11 dB/cm TM-polarized waveguide loss suggests some residual QD absorption at the probe wavelength, which may also yield a bleaching signal. A bleaching of the QD absorption results in increased probe transmission for both MZI arms, while a phase shift results in demodulated probe transmissions of opposite sign for both arms. We conclude that we observe predominantly all-optical switching, since the probe signals observed for the two arms are of opposite sign. A second argument for the observation of index of refraction nonlinearities are the strong oscillations of the demodulated probe transmission when we fine-tune the probe wavelength through the transmission characteristics of the MZI, which oscillate as a function of probe wavelength.

Finally, at 1150 nm pump wavelength when we also excite the waveguide core, we could clearly observe switching from the cross to the bar output of the MZI on an infrared camera. This confirms that we don't observe bleaching. At this excitation wavelength, a much larger fraction of the QDs is populated due to carrier capture from the waveguide core. An analysis of the data suggests that the observed switching at 1150 nm excitation wavelength is predominantly due to QD state-filling, while a small part might be due to InGaAsP bandfilling.

The pump-wavelength dependence of the all-optical switching signal is presented in Fig. 3. The slow decrease of the all-optical switching signal with increasing pump wavelength confirms that the signal does not arise from the exponentially decreasing Urbach tail in the InGaAsP. The all-optical switching thus is not due to residual absorption in the InGaAsP. At the excitation density applied, we also do not expect bandfilling in the InAs wetting layer. Summarizing, we conclusively interpret the observed all-optical switching as being due to state-filling in the InAs/InP QDs.

At $\lambda=1550$ nm, the pump laser is only resonant with the ground state of the largest InAs/InP QDs ($E_g \geq 1550$ nm), while excitation at $\lambda=1400$ nm is able to excite both the ground states of small QDs as well as the excited states of the larger QDs. The total QD absorption thus decreases with increasing wavelength, in accordance with our experimental observation.

The probe wavelength dependence of the all-optical switching signal is shown in Fig. 2. The switching efficiency is relatively wavelength insensitive due to the intentionally broad size distribution of the InAs/InP QDs. From the PL-spectrum, we observe that the QD PL varies less than 10% in the range 1470-1550 nm. A similar wavelength insensitivity is observed for the probe wavelength dependence.

We finally estimate the switching efficiency of the all-optical switch. From the results presented in Fig. 2, we observe $2.6 \cdot 10^{-4}$ rad phase shift at 0.125 mW pump power. Since 10% of this power directly excites the waveguide and the temporal duty cycle is 0.5%, we find a phase shift of 4.2 rad/mW incident power. We correct for the estimated $8 \cdot 10^{-4}$ absorption strength[9] of a single QD-layer with a QD-height of 7 nm, yielding a switching efficiency of 5 rad/(μW absorbed power) or an absorbed energy of 6 fJ for a π-phase change. The estimated index of refraction nonlinearity is $n_2$=0.08/(μW absorbed power). We present the nonlinearity as a function of the absorbed laser power, since this is the relevant quantity for all-optical switching, when the pump beam excites one arm of the MZI through a separate third waveguide[1,2]. We finally emphasize that the excellent switching efficiency is obtained from a 600 μm long phase shifter with a *single* QD-layer. The switching efficiency can further be enhanced by filling the waveguide core with multiple QD layers.

In conclusion, we observe all-optical switching in a Mach-Zehnder switch containing a single layer of QDs. The switching efficiency is 5 rad/(μW absorbed power). Due to the large size fluctuations of the InAs/InP QDs, we observe wavelength insensitive operation.

The pump wavelength dependence clearly shows that the all-optical switching is due to state-filling within the QDs. This work is part of the TUC project supported by the technology programme Towards Freeband Communication Impulse.

Figure Captions:

*Fig. 1*

*Schematic picture of the all-optical switching setup using a pump beam from the top to excite the QDs in the upper arm (in the shaded area) of the Mach Zehnder switch. The QDs are contained in the core of waveguides from which the switch is made.*

*Fig 2*

*Demodulated probe transmission of versus pump power, showing QD all-optical switching at a pump wavelength of 1450 nm and at probe wavelengths indicated in the figure.*

*Fig 3.*

*Pump wavelength dependence of the all-optical switching signal showing 2 measurement series. The inset shows the room temperature PL spectrum recorded at 256 mW/cm$^2$, showing the InAs/InP QDs luminescence at 1500 nm and the luminescence of the waveguide core at 1300 nm.*

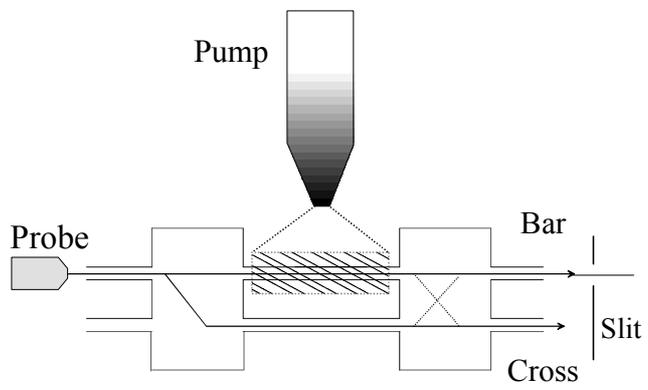

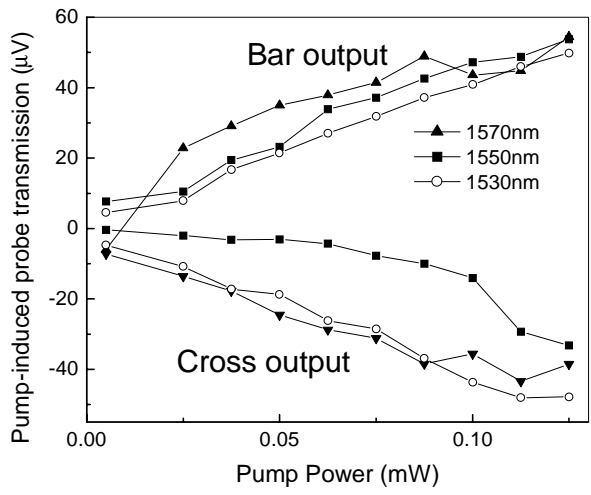

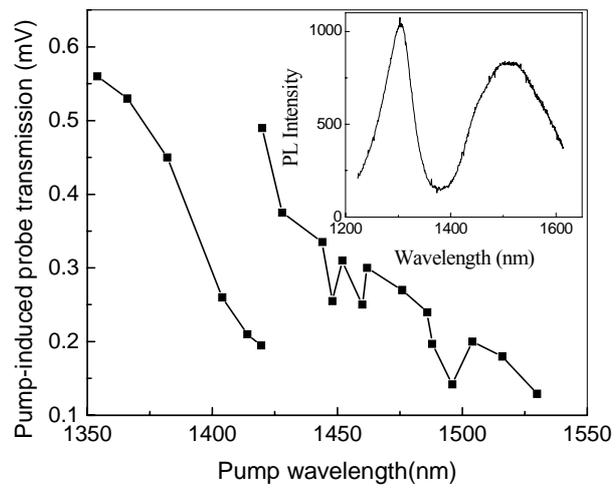